\documentclass[11pt]{article}
\usepackage{amssymb,latexsym}
\renewcommand{\em}{\sl}
\newtheorem{theorem}{Theorem}[section]
\newtheorem{proposition}[theorem]{Proposition}
\newtheorem{lemma}[theorem]{Lemma}

\newtheorem{definition}[theorem]{Definition}

\newcommand{\proof}{\noindent {\sc Proof.}}
\newcommand{\qed}{\rule{2mm}{4mm}}

\newcommand{\subf}{\subseteq^{\sf f}}

\newcommand{\Fin}[1]{{\sf Fin} ( #1 )}

\newcommand{\snd}{{\sf snd}}
\newcommand{\ltuple}{\mbox{${\langle}$}}
\newcommand{\rtuple}{\mbox{${\rangle}$}}

\newcommand{\restrict}[2]{\rho^{#1}_{#2}}

\newcommand{\ie}{{\em i.e.}}

\newcommand{\id}{{\sf id}}

\newcommand{\IFF}{\;\; \Longleftrightarrow \;\;}
\newcommand{\IFFDEF}{\;\; \stackrel{\rm def}{\Longleftrightarrow} \;\;}
\newcommand{\Iff}{\; \Leftrightarrow \;}
\newcommand{\implies}{\; \Rightarrow \;}
\newcommand{\Iffdef}{\IFFDEF}
\newcommand{\EQDEF}{\stackrel{\rm def}{=}}
\newcommand{\eqdef}{\EQDEF}
\newcommand{\AND}{\; \& \;}
\newcommand{\OR}{\; \mbox{or} \;}

\newcommand{\cref}{\vartriangleleft_{\alpha}}
\newcommand{\dmerge}{{\sf dmerge}}

\newcommand{\Cee}[1]{{\cal C}(#1 )}

\newcommand{\Tee}{{\cal T}}
\newcommand{\Vee}{{\cal V}}
\newcommand{\Ee}{{\cal E}}
\newcommand{\noo}[2]{\nu^{#1}_{#2}}
\newcommand{\Pee}[1]{{\Bbb P} #1}
\newcommand{\Ell}[1]{{\Bbb L} #1}
\newcommand{\Mod}{{\cal M}}
\newcommand{\ModL}{{\cal M}_{\Bbb L}}
\newcommand{\ModP}{{\cal M}_{\Bbb P}}
\newcommand{\netcomp}{\parallel}
\newcommand{\lfp}[1]{{\sf lfp} ( #1 )}
\newcommand{\Str}[1]{{\sf Str} ( #1 )}
\newcommand{\LTr}[1]{{\sf LTr}_{#1}}
\newcommand{\PTr}[1]{{\sf PTr}_{#1}}
\newcommand{\Con}{{\sf Con}}
\newcommand{\Chan}{{\sf Chan}}
\newcommand{\Sort}{{\bf Sort}}
\newcommand{\GKPL}{({\rm GKP}_l )}
\newcommand{\GKPS}{({\rm GKP}_s )}
\newcommand{\Cpo}{{\bf Cpo}}
\newcommand{\Cpos}{{{\bf Cpo}^s}}
\newcommand{\Alg}{\omega {\bf Alg}}
\newcommand{\Algs}{\omega {\bf Alg}^s}
\newcommand{\Incdom}{{\bf IncDom}}
\newcommand{\wcovers}{\preceq}
\newcommand{\covers}{\prec}
\newcommand{\relcover}[3]{[#1 , #2 ] \sqsubseteq #3}

\newcommand{\natarrow}{\stackrel{.}{\rightarrow}}

\newcommand{\converges}{\mbox{${\downarrow}$}}
\newcommand{\diverges}{\mbox{${\uparrow}$}}

\begin{document}
\title{A Generalized Kahn Principle for Abstract Asynchronous Networks}
\author{Samson Abramsky\\
\\
Department of Computing\\
Imperial College of Science, Technology and Medicine\\
180 Queen's Gate\\
London SW7 2BZ\\
England\\
\\
Published in the Proceedings of the Symposium on\\ 
Mathematical
Foundations of Programming Language Semantics,\\
Springer {\sl Lecture Notes in Computer Science} 442, pp. 1--21}
\date{November 3, 1989}
\maketitle
\begin{abstract}
Our general motivation is to answer the question: ``What is a model of
concurrent computation?''. As a preliminary exercise, we study 
dataflow networks. We develop a very general notion of model for
asynchronous networks. The ``Kahn Principle'', which states
that a network built from functional nodes is the least fixpoint of a system
of equations associated with the network, has become a benchmark for the 
formal study of dataflow networks.
We formulate a generalized version of the Kahn Principle, which applies to
a large class of non-deterministic systems, in the setting of abstract
asynchronous networks; and prove that the Kahn Principle holds under certain
natural assumptions on the model. We also show that a class of models,
which represent networks that compute over arbitrary event structures,
generalizing dataflow networks which compute over streams, satisfy 
these assumptions.
\end{abstract}
\bibliographystyle{alpha}
\section{Introduction}

There are by now a proliferation of mathematical structures which have
been proposed to model concurrent systems. These include synchronization
trees \cite{Win85a}, event structures \cite{Win86}, Petri nets \cite{Rei85}, 
failure sets
\cite{Hoa85}, trace monoids \cite{Maz89}, pomsets \cite{Pra82} and many others. One is then
led to ask: what general structural conditions should a model of
concurrency satisfy? There is an obvious analogy with the
$\lambda$-calculus, where a consensus on the appropriate notions of
model only emerged some time after a number of particular model
constructions had been discovered ({\sl cf.} \cite{Bar}). Indeed, we would
like to pose the question:
\begin{center}
``What is a model of concurrent computation?''
\end{center}
in the same spirit as the title of Meyer's excellent paper \cite{Mey82}.

One important disanalogy with the $\lambda$-calculus is that the field of
concurrent computation so far lacks a canonical syntax; and at a deeper
level, there is as yet no analogue of Church's thesis for concurrent
computation. The various formalisms which have been proposed actually
draw inspiration from a highly varied phenomenology: synchronous,
asynchronous, real-time, dataflow, shared-memory, declarative,
object-oriented, systolic, SIMD, neural nets, etc. etc. In these
circumstances, some more modest and circumscribed attempts at
synthesis seem justified. At the same time, merely finding general
definitions which subsume a number of concrete models is not enough;
good definitions should show their cutting edge by yielding some
non-trivial results.

In the present study, we start from a particular class of concurrent
systems, the {\em non-deterministic dataflow networks} \cite{Par82}. A
problem which has established itself as a benchmark for the formal study
of such systems is the {\em Kahn Principle} \cite{Kah74}, which states that if a
network is composed of functional nodes, its behaviour is captured by the
least fixpoint of a system of equations associated with the network in a
natural way.

We attempt to formulate a notion of model for such networks in the most
general and abstract form which still allows us to prove the Kahn
Principle. In this way, we hope both to shed light on the initial motivating
question of the axiomatics of process semantics, and to expose the
essence of the Kahn Principle. In the course of doing so, we shall attain a
level of generality, both as regards the notion of asynchronous network we
consider, and the statement of the Kahn Principle, far in excess of
anything we have seen in the literature.

The structure of the remainder of the paper is as follows. In section~2, we
review some background on domain theory and dataflow networks. Then in
section~3 we introduce our general notion of model, state a generalized
version of the Kahn Principle, and prove that certain conditions on models
are sufficient to imply the Kahn Principle. As far as I know, these are the
first results of this form, as opposed to proofs of the Kahn Principle for
specific models. Some directions for further research are given in
section~4.

\section{Background}

We begin with a review of some notions in Domain theory; see e.g.
\cite{GS89} for further information and motivation.

We write $\Fin{X}$ for the set of finite subsets of a set $X$; and $A \subf
X$ for the assertion that $A$ is a finite subset of $X$. A {\em poset} is a
structure $(P, {\leqslant})$, where $P$ is a set, and $\leqslant$ a reflexive,
transitive, anti-symmetric relation on $P$. Let $(P, {\leqslant})$ be a poset.
We write $\converges x = \{ y \in P \mid y \leqslant x \}$, $\diverges x = \{ y
\in P \mid y \geqslant x \}$ for $x \in P$; and $\converges X = \bigcup_{x \in X}
\converges x$, $\diverges X = \bigcap_{x \in X} \diverges x$ for 
$X \subseteq P$. A subset $S \subseteq P$ is {\em
directed} if every finite subset of $S$ has an upper bound in $S$. A poset is
{\em directed-complete} if every directed subset $S$ has a least upper
bound, written $\bigsqcup S$. A cpo (complete partial order) is a
directed-complete poset with a least element, written $\bot$. An element
$b \in D$ of a cpo $(D, {\sqsubseteq})$ is {\em compact} if whenever $S \subseteq D$ is
directed, and $b \sqsubseteq \bigsqcup S$, then $b \sqsubseteq d$ for
some $d \in S$. We write $K(D)$ for the set of compact elements of $D$,
and $K(d) = \converges d \cap K(D)$ for $d \in D$. A cpo $D$ is {\em
algebraic} if for all $d \in D$, $K(d)$ is directed, and $d = \bigsqcup K(d)$;
and $\omega$-algebraic if in addition $K(D)$ is countable. An {\em ideal}
over a poset $P$ is a directed subset $I \subseteq P$ such that $x \leqslant y
\in I \implies x \in I$. The {\em ideal completion} of a poset $P$ is the set
of ideals over $P$, ordered by inclusion. If $P$ has a least element, this is
an algebraic cpo; it is $\omega$-algebraic if $P$ is countable.

A map $f : D \rightarrow E$ of cpo's is {\em continuous} if for every
directed subset $S \subseteq D$, $f(\bigsqcup S) = \bigsqcup f(S)$; and
{\em strict} if $f(\bot_D ) = \bot_E$.
A subset $U \subseteq D$ of a cpo $D$ is {\em Scott-open} if $U = \diverges U$,
and whenever $\bigsqcup S \in U$ for a directed subset $S \subseteq D$,
then $S \cap U \not= \varnothing$. The Scott-open subsets form a topology on $D$;
a function between cpo's is continuous as defined above iff it is continuous
in the topological sense with respect to the Scott topology.
The Scott-open subsets of an algebraic cpo $D$ are those of the form
$\bigcup_{i \in I} \diverges b_i$, where $b_i \in K(D)$ for all $i \in I$.

We define some standard constructions on cpo's. Given a set $X$, the
algebraic cpo of {\em streams} over $X$, $\Str{X}$, is the set of finite and
infinite sequences over $X$, with the prefix ordering. If $D$, $E$ are cpo's,
$[D \rightarrow E]$ is the cpo of continuous functions from $D$ to $E$,
with the pointwise ordering; if $\{ D_i \}_{i \in I}$ is a family of cpo's,
$\prod_{i \in I} D_i$ is the cartesian product cpo, with the componentwise
ordering. If $f : D \rightarrow D$ is a continuous map on a cpo $D$, it has a
least fixed point, defined by
\[ \lfp{f} = \bigsqcup_{k \in \omega} f^k (\bot ) . \]

We shall assume some small knowledge of category theory in the sequel;
suitable references are \cite{Mac71,AM75}. We write $\Cpo$ for the category of
cpo's and continuous maps, $\Cpos$ for the subcategory of strict
continuous maps; and $\Alg$, $\Algs$ for the corresponding categories of
$\omega$-algebraic cpo's.

We define the {\em weak covering relation} on a poset $(P, {\leqslant})$ by:
\[ x \wcovers y \Iffdef x \leqslant y \AND \forall z. \, (x \leqslant z \leqslant y \implies
(x = z \OR y = z)) \]
and the {\em covering relation} by
\[ x \covers y \Iffdef x \wcovers y \AND x \not= y . \]
The computational intuition behind the covering relation as used in Domain
theory is that it represents an atomic computation step, or the occurrence
of an atomic event; this idea can be traced back to \cite{KP78}.

A {\em covering sequence} in an algebraic cpo $D$ is a non-empty finite or
infinite sequence of compact elements $(b_n )$, such that $b_0 = \bot$,
and $b_n \covers b_{n+1}$ for all terms $b_n$, $b_{n+1}$ in the sequence.
A covering sequence can be taken as a representation of $d = \bigsqcup
b_n$, which gives a step-by-step description of how it was computed. 

Given an algebraic cpo $D$, we can form the algebraic cpo ${\cal C}(D)$ of
covering sequences over $D$, with the prefix ordering. There is a
continuous map $\mu : {\cal C}(D) \rightarrow D$, with $\mu ((b_n )) =
\bigsqcup b_n$. 

Finally, we define the {\em relative covering relation} in $D$ by:
\[ \relcover{b}{c}{d} \Iffdef b, c \in K(d) \AND b \covers c. \]
We can think of $b \covers c$ as an atomic step at some finite stage in the
computation of $d$.

A {\em prime event structure} \cite{Win86} is a structure $\Ee = (E,
{\leqslant}, \Con)$, where $(E, {\leqslant})$ is a countable poset, and $\Con \subseteq
\Fin{E}$ a family of finite subsets of $E$, satisfying: 

\begin{itemize}

\item $\forall e \in E. \, ( \converges e \;\; \mbox{is finite})$.

\item $\forall e \in E. \, ( \{ e \} \in \Con )$.

\item $A \subseteq B \in \Con \implies A \in \Con$.

\item $A \in \Con \implies \converges A \in \Con$.

\end{itemize}

We refer to elements of $E$ as {\em events}, to $\leqslant$ as the {\em
causality} or {\em enabling} relation, and to $\Con$ as the {\em
consistency} predicate. A {\em configuration} of $\Ee$ is a set $x
\subseteq E$ such that

\begin{itemize}

\item $e \leqslant e' \in x \implies e \in x$

\item $A \subf x \implies A \in \Con$.

\end{itemize}

The set $| \Ee |$ of configurations of $\Ee$, ordered by inclusion, is an
algebraic cpo; the compact elements are the finite configurations. Note
that in $| \Ee |$, $x \covers y$ iff $y \setminus x = \{ e \}$ for some $e \in
E$; and that if $x \sqsubseteq y$ for compact elements $x$, $y$, there is a
sequence $e_1 , \ldots , e_n$ such that $x = z_0 \covers \cdots \covers
z_n = y$, where $z_i = x \cup \{ e_1 , \ldots , e_i \}$. The algebraic cpo's
which arise from prime event structures are characterized in \cite{Win86}; we
refer to them as {\em event domains}. They form quite an extensive class,
containing models of type-free and polymorphic lambda calculi (using
stable functions), as well as the usual datatypes of functional
programming \cite{CGW87}.

We now turn to the dataflow model of concurrent computation. Consider a
process network, represented by a directed (multi)graph $G = (N, A, s, t)$,
where $N$ is the set of nodes, $A$ the set of arcs, and $s, t : A
\rightarrow N$ are the source and target functions. Each node is labelled
with a sequential process, while each arc corresponds to a buffered
message channel, which behaves like an unbounded FIFO queue. In addition
to the usual sequential constructs, each node $n$ can read from its input
channels (those $\alpha$ with $t(\alpha ) = n$), and write to its output
channels (those $\alpha$ with $s(\alpha ) = n$). Although this
computational model might be criticised as unrealistic because of the
unbounded buffering, this very feature enables a high degree of
parallelism, and the model is appealingly simple, and quite close to a
number of actually proposed and implemented dataflow languages and
architectures \cite{AW85,KLP79,KM77,GGKW84}. Kahn's brilliant insight in his seminal paper
\cite{Kah74} was that the behaviour of such networks could be captured
denotationally in a very simple and elegant fashion, using some elementary
domain theory. The key idea is to model the behaviour of each message
channel $\alpha$, on which atomic values from the set $D_{\alpha}$ can be
transmitted, as a stream from the domain $\Str{D_{\alpha}}$. Using
standard denotational techniques, the behaviour of the process at node $n$,
with input channels $\alpha_1 , \ldots , \alpha_k$, and output channnels
$\beta_1 , \ldots , \beta_l$, can be modelled by a continuous function
\[ f : \Str{D_{\alpha_1}} \times \cdots \times \Str{D_{\alpha_k}}
\rightarrow \Str{D_{\beta_1}} \times \cdots \times \Str{D_{\beta_l}} .
\]
The behaviour of the whole system can be modelled by setting up a system
of equations, one for each channel in the network, of the overall form
\[ {\bf X} = G({\bf X}), \]
where $G : \prod_{\alpha} \Str{D_{\alpha}} \rightarrow  \prod_{\alpha}
\Str{D_{\alpha}}$; and solving by taking the least fixed point $\lfp{G} \in
\prod_{\alpha} \Str{D_{\alpha}}$.

It is worth noting that Kahn never {\em proved} the coincidence of this
denotational semantics with an operational semantics based directly on
the computational model sketched above; indeed, he never defined any
formal operational semantics for dataflow networks. Nevertheless, no-one
has ever seriously doubted the accuracy of his semantics. A number of
subsequent attempts have been made to fill this gap in the theory \cite{Fau82,LS88};
it has proved surprisingly difficult to give a clean and elegant account.

In another direction, many attempts have been made to overcome one
crucial limitation built into Kahn's framework; namely, the assumption
that all processes in the network are deterministic, and hence their
behaviour can be described by functions. This limitation must be overcome
in order for these networks to be sufficiently expressive to model
general-purpose concurrent systems (see e.g. \cite{Hen82,Abr84}). However, as soon as
non-deterministic processes are allowed, the denotational description of
dataflow networks becomes much more complicated. In fact,  naive
attempts to extend Kahn's model have been shown to be doomed to failure
by certain ``anomalies'' which were found by Keller \cite{Kel78} and Brock and
Ackerman \cite{BA81}. In particular, Brock and Ackerman exhibited a pair of
deterministic processes $N_1$, $N_2$ with the same Kahn semantics, and
a non-deterministic context $C[\cdot ]$ such that $C[N_1 ] \not= C[N_2 ]$ with
respect to the intended operational semantics. The main point of this is 
to show that in the presence
of non-determinism, the behaviour of a system is no longer adequately
modelled by a ``history tuple'' $d \in \prod_{\alpha} \Str{D_{\alpha}}$.
Such a tuple records the order in which values are realized on each
channel, but fails to record causality relations which may exist between
items of data on {\em different} channels. A number of more detailed
models have  been proposed which reflect this kind of information. Two in
particular have received some attention.

\begin{definition}
Let $S$ be a set of channel names, where for each $\alpha \in S$, there is
a set $D_{\alpha}$ of atomic data which can be transmitted over $\alpha$.
The domain of {\em linear traces} over $S$, $\LTr{S}$, is the stream domain
$\Str{E_S}$, where
\[ E_S = \{ (\alpha , d) \mid \alpha \in S, d \in D_{\alpha} \} . \]
\end{definition}

The idea is that a linear trace represents a sequential observer's view of a
computation in the network, as a sequence of atomic events $(\alpha ,
d)$---namely, the production of the atomic value $d$ on the channel
$\alpha$. We can regard linear traces as more detailed---perhaps even
{\em over}-specified---representations of history tuples; indeed, there is
an obvious ``result'' or ``output'' map $\mu_S : \LTr{S} \rightarrow
\prod_{\alpha \in S} \Str{D_{\alpha}}$. It is a useful exercise to verify
that this is strict and continuous.

Given $S \supseteq T$, we can define a (strict, continuous) {\em
restriction map}, $\restrict{S}{T} : \LTr{S} \rightarrow \LTr{T}$, where
$\restrict{S}{T}(s)$ is obtained by deleting all $(\alpha , d)$ from $s$ such
that $\alpha \not\in T$.

In the linear trace model, a process is modelled by a pair $(S, P)$, where
$S$ is the set of channels incident to the process, and $P \subseteq
\LTr{S}$ describes its (possibly non-deterministic) behaviour. The key
definition is that of the operation of {\em network composition}, which
glues together a family of processes along their coincident channels. Let
$\{ (S_j , P_j )\}_{j \in J}$ be a family of processes; we define
$\parallel_{j \in J} (S_j , P_j ) = (S, P)$, where
\[ \begin{array}{rcl}
S & = & \bigcup_{j \in J} S_j \\
P & = & \{ s \in \LTr{S} \mid \forall j \in J. \, (\restrict{S}{S_j}(s) \in P_j ) \}
.
\end{array} \]

Note that this definition of the behaviour of a net is quite different in
form to the Kahn semantics; we have replaced continuous functions by sets
of traces, and the iterative construction of a least fixed point by a
product-like construction. It thus becomes a matter of some importance to
see if this definition actually {\em coincides} with the Kahn semantics in
the case when each node in the network is in fact computing some
continuous function. (Of course, we must firstly define what that means in
terms of sets of traces). We refer to this task as the proof of the {\em
Kahn Principle} for the linear trace model.

The linear trace model has recently been proved to be {\em fully abstract}
in a certain sense \cite{Jon89}; however, some other models have also received
considerable attention, and avoid the apparent over-specification of linear
traces. In particular there are the {\em pomset} models \cite{Pra82}, which
were inspired by Brock and Ackerman's {\em scenarios} \cite{BA81}. The idea is to
allow partial orders of events, rather than insisting on purely sequential
observations.

\begin{definition}
The domain of partially ordered traces $\PTr{S}$ is the ideal completion of the
finite partially-ordered traces with the prefix ordering, where:

\begin{itemize}

\item A finite partially-ordered trace is an isomorphism type of finite
labelled partial orders $(V, {\leqslant}, \ell )$, where $\ell : V \rightarrow
E_S$, and for each $\alpha \in S$, the subposet 
\[ \{ v \in V \mid \exists d \in D_{\alpha} (\ell (v) = (\alpha , d)) \} \]
is linearly ordered.

\item The prefix ordering is defined on representatives by:
\[ \begin{array}{rcl}
(V, {\leqslant}, \ell ) \sqsubseteq (V', {\leqslant}', \ell' ) & \IFFDEF & V \subseteq V'
\AND {\leqslant} = {\leqslant}' \cap V^2 \AND \ell = \ell' \restriction V \\
& &  \AND v \leqslant' v' \in V \implies v \in V. 
\end{array} \]
\end{itemize}
\end{definition}
Note that, if we identify sequences with isomorphism types of labelled 
{\em linear} orders, we have the inclusion $\LTr{S} \subseteq \PTr{S}$.
Once again, there is an evident definition of a restriction map
$\restrict{S}{T} : \PTr{S} \rightarrow \PTr{T}$ for $S \supseteq T$, and,
by virtue of the stipulation that events at each channel are linearly
ordered, a map $\mu_S : \PTr{S} \rightarrow \prod_{\alpha \in S}
\Str{D_{\alpha}}$.

We can then define the notion of network composition in the partially
ordered trace model in {\em exactly the same way} as we did for the linear
traces, modulo the different notions of ``trace'' and ``restriction''; and
formulate the Kahn Principle in exactly the same terms.
The main previous work on proving the Kahn Principle for (essentially) the
partially ordered trace model is described in \cite{GP87}.

Our aim is firstly to extract the essential properties of this situation to
arrive at a general notion of model, and then to prove the Kahn principle in
this general setting. Apart from yielding the particular results for the
linear and partially-ordered trace models for dataflow networks as
instances of our general result, there are a number of other insights that
we hope this work provides:

\begin{itemize}

\item The abstract networks we consider 
compute over  a much broader class
of domains than just the stream domains of dataflow---our results apply 
at least to the event domains.

\item The version of the Kahn Principle we formulate and prove in fact
applies not only to the deterministic case, but to a broad class of {\em
non-deterministic} networks---namely those in which each node computes
one of a {\em set} of possible continuous functions. This includes for
example the so-called ``infinity-fair merge'', though not the ``angelic
merge'' \cite{PS88}. As far as I know, this major extension to the Kahn
Principle is new, even for the specific models described above.

\item Although our notion of model is abstracted from the dataflow
family, and cannot be claimed to be fully general, we hope it is a useful
step along the way to answering the question raised in the opening paragraph, namely: ``what is a model of concurrent computation?''.
\end{itemize}
\section{Results}

\subsection{Models}

We assume a class $\Chan$ of {\em channel names}, ranged over by
$\alpha$, $\beta$, $\gamma$. We refer to sets of channels as {\em sorts};
the class of sorts, partially ordered by inclusion, is denoted by $\Sort$.
We use $S$, $T$, $U$ to range over sorts.

\begin{definition}

A {\em model} $\Mod = (\Tee , \Vee , \mu )$ comprises:

\begin{itemize}

\item functors $\Tee , \Vee : \Sort^{\sf op} \rightarrow \Cpos$

\item a natural transformation $\mu : \Tee \natarrow \Vee$

\end{itemize}

such that $\Vee$ preserves limits.

\end{definition} 

We refer to $\Tee_S$ as the {\em traces} of sort $S$, $\Vee_S$ as the {\em
values} of sort $S$, and $\mu$ as the {\em output} or {\em evaluation}
map.

More explicitly, $\Tee$ assigns to each sort $S$ a cpo $\Tee_S$, and to
each $S \supseteq T$ a strict, continuous {\em restriction map}
$\restrict{S}{T} : \Tee_S \rightarrow \Tee_T$, such that:

\begin{itemize}

\item $S \supseteq T \supseteq U \implies \restrict{T}{U} \circ
\restrict{S}{T} = \restrict{S}{U}$

\item $\restrict{S}{S} = \id_{\Tee_S}$.

\end{itemize}

Similarly, $\Vee$ assigns a cpo $\Vee_S$ to each sort $S$. The
requirement that $\Vee$ preserves limits amounts to asking that $\Vee$
takes {\em unions} in $\Sort$ to {\em products} in $\Cpos$. Since each
sort is the union of its singletons, this means that if $\Vee_{\alpha}$ is
the value domain of sort $\{\alpha\}$,
\[ \Vee_S = \prod_{\alpha \in S} \Vee_{\alpha} ; \]
and that the restriction maps will be the projections onto sub-products:
for $S \supseteq T$, $\pi^{S}_{T} : \Vee_S \rightarrow \Vee_T$. Thus
$\Vee$ is completely determined by the $\Vee_{\alpha}$.

Finally, for each sort $S$ there is a strict, continuous map $\mu_S :
\Tee_S \rightarrow \Vee_S$, such that for all $S \supseteq T$,
\[ \mu_T \circ \restrict{S}{T} = \pi^S_T \circ \mu_S . \]
{\bf Notation.} We write $\noo{S}{T} = \mu_T \circ \restrict{S}{T} =
\pi^S_T \circ \mu_S$.

\subsubsection*{Examples}

(1). Firstly, from the discussion in the previous Section, it is easy to see
that both linear and partially-ordered traces yield examples of models.
More precisely, for each channel $\alpha$ fix a set $D_{\alpha}$ of atomic
values; then define $\Vee_{\alpha} = \Str{D_{\alpha}}$, and $\Tee_S =
\PTr{S}$ $(\LTr{S})$, $\restrict{S}{T}$, $\mu_S$ as in Section~2. The
verification of the required functoriality and naturality conditions is
straightforward.

\noindent (2). We now describe a general class of models. For each channel $\alpha$,
fix an event structure $\Ee_{\alpha} = (E_{\alpha}, {\leqslant}_{\alpha},
\Con_{\alpha})$. Define $\Vee_{\alpha} = | \Ee_{\alpha} |$, the domain of
configurations over $\Ee_{\alpha}$. For a sort $S$, we define $\Ee_S =
\prod_{\alpha \in S} \Ee_{\alpha}$, where the product of event structures
is defined as their {\em disjoint union} \cite{Win86}: $\Ee_S = (E_S, {\leqslant}_S,
\Con_S )$, where 
\[ \begin{array}{rcl}
E_S & \eqdef & \{ (\alpha , e) \mid \alpha \in S, e \in E_{\alpha} \} \\
(\alpha , e) \leqslant_S (\beta , e' ) & \Iffdef & \alpha = \beta \AND e
\leqslant_{\alpha} e' \\
A \in \Con_S & \Iffdef & \forall \alpha \in S. \, ( \{ e \mid (\alpha , e) \in A
\} \in \Con_{\alpha} ) .
\end{array} \]

We have \cite{Win86}: $| \Ee_S | \cong \prod_{\alpha \in S} | \Ee_{\alpha} |$,
and we shall take $\Vee_S = | \Ee_S |$. For $S \supseteq T$, the
projections $\pi^S_T : | \Ee_S | \rightarrow | \Ee_T |$ are defined by
$\pi^S_T (x) = x \cap E_T$.

In order to define the traces over $\Ee_S$, we follow the idea that

\begin{center}
traces = data + causality.
\end{center}

Thus a trace is a configuration together with extra information about the
order in which data was actually produced in a particular computation,
reflecting some causal constraints.

\begin{definition}

A {\em trace} over an event structure $\Ee = (E, {\leqslant}, \Con )$ is a pair
$t = (x_t , \leqslant_t )$, where $x_t \in | \Ee |$, and $\leqslant_t$ is a partial
order on $x_t$ such that:

\begin{itemize}

\item $\forall e \in x_t . \, ( \{ e' \in x_t \mid e' \leqslant_t e \} \;\; \mbox{is finite}
)$

\item $({\leqslant} \cap x_t^2 ) \subseteq {\leqslant}_t$. 

\end{itemize}

Traces are partially ordered as follows:

\[ t \sqsubseteq t' \Iffdef x_t \subseteq x_{t'} \AND {\leqslant}_t = {\leqslant}_{t'}
\cap x_t^2 \AND (e \leqslant_{t'} e' \in x_t \implies e \in x_t ). \]

\end{definition}

Clearly, traces with this ordering form an algebraic cpo $\Pee{\Ee}$. A
trace $t$ is {\em linear} if $\leqslant_t$ is a linear order; the linear traces
also form an algebraic cpo, $\Ell{\Ee}$, and $\Ell{\Ee} \subseteq
\Pee{\Ee}$. 
The compact elements of $\Pee{\Ee}$ are those $t$ for which $x_t$ is
a finite configuration of $| \Ee |$. Also, $t \covers u$ in $\Pee{\Ee}$
iff $x_u \setminus x_t = \{ e \}$ for some $e$ which is {\em maximal} in 
$\leqslant_u$.
The following construction on trace domains will be useful. Given
$t \in \Pee{\Ee}$, and $X \subseteq x_t$, we define $t {\restriction} X$ by:
\[ \begin{array}{rcl}
x_{t {\restriction} X} & = & \{ e \in x_t \mid \exists e' \in X. \, e \leqslant_t e' \} \\
\leqslant_{t {\restriction} X} & = & \leqslant_t \cap (x_{t {\restriction} X})^2 .
\end{array} \]
Clearly $t {\restriction} X$ is a well-defined trace, and $t {\restriction} X \sqsubseteq t$; moreover, $X \subseteq Y \implies t {\restriction} x 
\sqsubseteq t {\restriction} Y$. 
This construction can also be applied to $\Ell{\Ee}$.

We can now complete the definitions for our two families of models,
$\ModP$ (partially ordered traces over event structures) and $\ModL$ (the
sub-model of linearly ordered traces). The trace domains for $\ModP$ are
defined by $\Tee_S = \Pee{\Ee_S}$, and for $\ModL$ by $\Tee_S =
\Ell{\Ee_S}$. The evaluation maps are defined for both by
\[ \mu_S (t) = x_t , \]
and the restriction maps by
\[ \restrict{S}{T} (t) = (x_t \cap E_T , {\leqslant}_t \cap E_T^2 ) , \]
for $S \supseteq T$.

The verification that these definitions yield models is straightforward.
Note that $\ModP$ and $\ModL$ are really {\em families} of models,
parameterized by the choice of event structures $\Ee_{\alpha}$ for each
$\alpha$. Our results will apply to {\em all} models in these families.

We now show how the concrete dataflow models of (1) are special cases of
$\ModP$ and $\ModL$. Fix a set $D_{\alpha}$ for each channel $\alpha$, and
define an event structure $\Ee_{\alpha}$ as follows:

\begin{itemize}

\item $E_{\alpha} = \{ (s, sd) \mid s \in D_{\alpha}^{\star}, d \in
D_{\alpha} \}$.

\item $(s, sd) \leqslant_{\alpha} (s' , s'd') \Iffdef sd \sqsubseteq s'd'$.

\item $A \in \Con  \Iffdef \forall (s, sd), (s' , s'd' ) \in A. ( sd \sqsubseteq s'd'
\OR s'd' \sqsubseteq sd)$.

\end{itemize}

It can easily be verified that $| \Ee_{\alpha} | \cong \Str{D_{\alpha}}$.
Also, we have

\begin{proposition}

For all sorts $S$,

\[ \begin{array}{rcl}

\PTr{S} & \cong & \Pee{\Ee_S} \\

\LTr{S} & \cong & \Ell{\Ee_S} .

\end{array} \]

\end{proposition}

\proof\ Given $t \in K(\Pee{\Ee_S})$, we define a labelled poset
$(x_t , {\leqslant}_t , \ell )$, where 
\[ \ell ((\alpha , (s, sd))) = (\alpha , d). \]
This defines a map $\phi : K(\Pee{\Ee_S}) \rightarrow K(\PTr{S})$.
(Note that the condition $(\leqslant_S \cap x_t ) \subseteq \leqslant_t$ is needed
to ensure that $\alpha$-events are linearly ordered in $\phi (t)$ for each
$\alpha \in S$).
Now consider a trace in $K(\PTr{S})$ with representative labelled poset
$(V, {\leqslant}, \ell )$. For each $v \in V$, let $\ell (v) = (\alpha , d)$.
The set of $\alpha$-labelled predecessors of $v$ is linearly ordered,
say 
\[ v_1 < \cdots < v_n < v, \]
and hence yields a finite sequence
$s = d_1 \cdots d_n \in K(\Str{D_{\alpha}})$, where $d_i = \snd \circ \ell (v_i )$, $i = 1, \ldots , n$.
We can thus define a new labelling function $\ell'$, which maps $v$ to
$(\alpha , (s, sd)) \in E_S$. Note that $\ell'$ is {\em injective}, and hence
we can dispense with $V$, and take the induced order on $\ell' (V)$:
$\ell' (v) \leqslant' \ell' (v' ) \Iffdef v \leqslant v'$, yielding a trace
$(\ell' (V), {\leqslant'})$ in $\Pee{\Ee_S}$.
Thus we obtain a map $\psi : K (\PTr{S}) \rightarrow K(\Pee{\Ee_S})$.
It is easily checked that $\phi$ and $\psi$ are monotone and mutually
inverse, yielding an order-isomorphism $K(\Pee{\Ee_S}) \cong K(\PTr{S})$,
and hence by algebraicity, $\Pee{\Ee_S} \cong \PTr{S}$. Finally, $\phi$,
$\psi$ cut down to an isomorphism $K(\Ell{\Ee_S}) \cong K(\LTr{S})$, and so
$\Ell{\Ee_S} \cong \LTr{S}$. \qed

One further connection will be useful: the linear traces over an event
structure are isomorphic to the covering sequences over its domain of
configurations.

\begin{proposition}
\label{covseqprop}
For any event structure $\Ee$, $\Ell{\Ee} \cong \Cee{| \Ee |}$.

\end{proposition}

\proof\ From our description of covering relations in event domains,
it follows that any covering sequence in $| \Ee |$ has the form
\[ x_0 \covers \cdots x_n \covers \cdots \]
where $x_0 = \varnothing$, $x_{n+1} \setminus x_n = \{ e_n \}$ for some
$e \in E$. We can then define the linear trace $t$ with $x_t = \bigcup x_n$,
$e_n \leqslant_t e_m \Iff n \leqslant m$. Conversely, any linear trace must,
by countability of $E$ and the well-foundedness property of traces, amount to a 
(finite or infinite) sequence $(e_n )$, from which we can define a
covering sequence $(x_n )$, where $x_n = \{ e_j \mid j \leqslant n \}$.
The fact that each $x_n \in | \Ee |$ follows from the conditions on traces.
These passages between $\Ell{\Ee}$ and $\Cee{\Ee}$ are easily checked to be
monotone and mutually inverse, establishing the required isomorphism. \qed
 
For the remainder of this section, we fix a model $\Mod = (\Tee , \Vee ,
\mu )$.

\begin{definition}

A {\em process} in $\Mod$ is a pair $(S, P)$, where $P \subseteq \Tee_S$.
Let $\{ (S_j , P_j ) \}_{j \in J}$ be a family of processes. The {\em network
composition} of this family is defined by:
\[ \netcomp_{j \in J} (S_j , P_j ) = (S, P), \]
where
\[ \begin{array}{rcl}
S & = & \bigcup_{j \in J} S_j \\
P & = & \{ t \in \Tee_S \mid \forall j \in J. \, (\restrict{S}{S_j} (t) \in P_j )
\} . 
\end{array} \]

\end{definition}
This definition was predictable from our discussion of concrete dataflow
models in the previous section. The next definition is a key one, which
answers the question of how to characterize when a process, {\em qua}
set of traces, is computing a function. In fact, we deal with the more
general situation when a process is computing any one (non-deterministically
chosen) from a {\em set} of functions.
\begin{definition}

Let $(S, P)$ be a process, with $S = I \cup O$, and let $F \subseteq [\Vee_I
\rightarrow \Vee_O ]$ be a set of continuous functions. We say that $(S,
P)$ {\em computes} $F$ if for all $t \in \Tee_S$:
\[ \begin{array}{l}
t \in P \IFF \exists f \in F: \\
(1) \;\; \noo{S}{O}(t) = f(\noo{S}{I}(t)) \\
(2) \;\; \relcover{u}{v}{t} \implies \noo{S}{O}(v) \sqsubseteq
f(\noo{S}{I}(u)).
\end{array} \]

\end{definition}
Condition (1) in this definition is the obvious stipulation that the overall
effect of the trace is to compute an input-output pair in the graph of
one of the functions $f \in F$. Condition (2) is more subtle; it insists
that the {\em way} this input-output pair is computed must be ``causally
consistent'', in the sense that for any step $u \covers v$
towards computing $t$, the output values realized after the step---at $v$---are
no more than what was justified as $f$ applied to the input values available
before the step---at $u$.\footnote{These conditions were directly inspired by Misra's
``limit'' and ``smoothness'' conditions in his notion of {\em descriptions}
\cite{Mis89}; his definition was made in the specific setting of the linear
trace domain $\LTr{S}$, and in a rather different context.}

As regards the generality conferred by the use of {\em sets} of functions,
consider the following example from dataflow \cite{Par82}: the deterministic merge
function
\[ \dmerge : \Str{X} \times \Str{X} \times \Str{\{ 0, 1 \}} \rightarrow \Str{X} \]
which uses an oracle to guide its choices. This satisfies the equations:
\[ \begin{array}{rcl}
\dmerge (a : x, y, 0:o ) & = & a : \dmerge (x, y, o) \\
\dmerge (x, b : y, 1 : o) & = & b : \dmerge (x, y, o) .
\end{array} \]
Now for any set of oracles $O$ we can define:
\[ F = \{ \lambda x, y. \dmerge(x, y, o) \mid o \in O \} . \]
If we take $O$ to be the set of {\em fair} oracles, \ie\  infinite binary sequences
containing infinitely many zeroes and infinitely many ones, then $F$ corresponds
to the ``infinity-fair merge'' \cite{PS88}; however, note that the ``angelic
merge'' cannot be obtained in this way.

Now let $\{ (S_j , P_j ) \}_{j \in J}$ be a family of processes, with $(S, P)
= \netcomp_{j \in J} (S_j , P_j )$. We say that $\{ (S_j , P_j ) \}_{j \in J}$
is a {\em non-deterministic functional network} if the following
conditions hold:

\begin{enumerate}

\item For all $j \in J$, $S_j = I_j \cup O_j$ and $(S_j , P_j )$ computes
$F_j \subseteq [\Vee_{I_j} \rightarrow \Vee_{O_j}]$.

\item For all $\alpha \in S$, there is exactly one $j \in J$ with $\alpha
\in O_j$.

\end{enumerate}

If $F_j$ is a singleton for all $j \in J$, we say that the network is {\em
deterministic}.

Condition (2) is worth some comment. The constraint that each channel has
{\em at most} one producer precludes non-determinism by ``short circuit''.
The requirement that there be {\em exactly} one producer is a technical
convenience; it means that we can avoid considering input channels---\ie\  
those
with no producer in the system---separately. 
Of course, we can still handle input channels, in a ``pointwise'' fashion;
for each given input value, we add a process which behaves like the
constant function producing that value on the channel. Indeed, in our
approach this is immediately generalized to allow a {\em set} of values
to be produced.

Now we generalize the Kahn semantics for dataflow in the obvious way.
For each $f \in \prod_{j \in J} F_j$, we define $G_f : \Vee_S \rightarrow
\Vee_S$ by:
\[ G_f = \ltuple \pi^{O_j}_{\alpha} \circ f_j \circ \pi^S_{I_j} \rtuple_{\alpha
\in S, \alpha \in O_j} . \]

By virtue of condition (2) on the network, there is exactly one component
of the tuple defining $G_f$ for each $\alpha \in S$.

We say that the network satisfies the {\em Generalized Kahn Principle} if
the following condition holds:
\[ \mbox{(GKP)} \quad \mu_S (P) = \{ \lfp{G_f} \mid f \in \prod_{j \in J}
F_j \} . \]
We say that $\Mod$ satisfies the Generalized Kahn Principle if (GKP) holds
for every non-deterministic functional network in $\Mod$. We say that
$\Mod$ satisfies the (ordinary) Kahn Principle if (GKP) holds for every
deterministic functional network. Note that in this case, $\prod_{j \in J}
F_j$ is a singleton, and hence so is the right-hand side of (GKP).

Our main objective will be to give sufficient conditions on $\Mod$ to
ensure that (GKP) holds. (GKP) states an equality between two sets; it is
convenient to consider the two inclusions separately. Firstly, we have
\[ \GKPS \quad \mu_S (P) \subseteq \{ \lfp{G_f} \mid f \in \prod_{j \in J}
F_j \} . \]
This is a {\em safety} property, since it asserts that every behaviour of
the network computes one of the values specified by the (generalized)
Kahn semantics. The converse:
\[ \GKPL \quad \{ \lfp{G_f} \mid f \in \prod_{j \in J} F_j \} \subseteq
\mu_S (P)  \]
is a {\em liveness} property, since it asserts that every specified value is realized by some computation.

\subsection{Safety}

\begin{definition}

An $\omega$-algebraic cpo is {\em incremental} if whenever $b
\sqsubseteq c$ in $K(D)$, there is a finite covering sequence
\[ b = b_0 \covers \cdots \covers b_n = c . \]
A strict, continuous function $f : D \rightarrow E$ on incremental domains
is an {\em incremental morphism}
if:

\begin{itemize}

\item $f$ weakly preserves relative covers:
\[ \relcover{b}{c}{d} \implies \relcover{f(b)}{f(c)}{f(d)} \OR f(b) = f(c) \in
K(d) . \]

\item $f$ lifts relative covers:
\[ \relcover{b'}{c'}{d'} = f(d) \implies \exists b, c. (\relcover{b}{c}{d} \AND
f(b) = b', f(c) = c' ). \]

\end{itemize}

\end{definition}

Incremental domains and morphisms form a category $\Incdom$. We say
that a functor $F : {\Bbb C} \rightarrow \Cpos$ is incremental if it factors
through the inclusion $\Incdom \hookrightarrow \Cpos$, and that a model
$\Mod = (\Tee , \Vee , \mu )$ is incremental if $\Tee$ is.

Note that all event domains, and all ideal completions of countable posets
satisfying both the ascending and descending chain conditions, are
incremental. The reason for our terminology is that incremental domains
are precisely the specialization to posets of the incremental categories
introduced in \cite{GJ88}.

\begin{proposition}

\label{modprop1}

$\ModP$ and $\ModL$ are incremental.

\end{proposition}

\proof\ We have already observed that the domains $\Pee{\Ee_S}$,
$\Ell{\Ee_S}$ are incremental. The fact the restriction maps weakly
preserve relative covers follows easily from the definitions. We must
verify the lifting property. We give the argument for $\ModP$ only.
Suppose then that $S \supseteq T$, $\relcover{{u'}}{v'}{t'}$ in $\Pee{\Ee_T}$,
and $\restrict{S}{T}(t) = t'$. We define $v = t \restriction x_{v'}$.
Since $x_{v'} \subseteq x_{t'} \subseteq x_t$, this is well-defined, 
and yields
$v \sqsubseteq t$. Let $w = \restrict{S}{T}(v)$. Since $x_{v'} \subseteq x_v$,
$x_{v'} \subseteq x_w$. For the converse, suppose $e \in x_w$. 
This implies that
$e \in E_T$, and that for some $e' \in x_{v'}$, $e \leqslant_t e'$. 
But this implies $e \leqslant_{t'} e'$, since $\restrict{S}{T}(t) = t'$, and
hence $e \in x_{{v'}}$, since ${v'} \sqsubseteq t'$ and $e' \in x_{v'}$. Thus 
$x_w = x_{v'}$. The same reasoning shows that $\leqslant_w = \leqslant_{v'}$,
and so $w = {v'}$.

To define $u$, recall that ${u'} \covers {v'}$ iff $x_{v'} \setminus x_{u'} = \{ e \}$
for some $e \in E_T$ which is maximal in $\leqslant_{v'}$. But then $e$ must also
be maximal with respect to $\leqslant_v$, since otherwise we would have
$e <_v e' \in x_{v'}$, which would imply $e <_{v'} e'$, contradicting 
$<_{v'}$-maximality of $e$. Thus if we define $x_u = x_v \setminus \{ e \}$,
$\leqslant_u = \leqslant_v \cap x_u^2$, we see that $v \restriction x_u = (x_u, \leqslant_u )$. Clearly $u \covers v$; and if $w = \restrict{S}{T}(u)$,
\[ x_w = x_u \cap E_T = (x_v \setminus \{ e \} ) \cap E_T = (x_v \cap E_T ) \setminus \{ e \} = x_{v'} \setminus \{ e \} = x_{u'} . \]
Similarly $\leqslant_w = \leqslant_{u'}$, yielding $\restrict{S}{T}(u) = {u'}$,
and the proof is complete. \qed

Our main objective in the remainder of this subsection is to prove:

\begin{theorem}

\label{th1}

If $\Mod$ is incremental, it satisfies $\GKPS$.

\end{theorem}
Our strategy is to use incrementality of the restriction maps to move
between local conditions expressing the functional behaviour of the nodes
and global conditions expressing the functional behaviour of the whole
network.

\begin{lemma}

\label{lem1}

Let $(S, P)$ be a non-deterministic functional process computing $F$,
where $S = I \cup O$. For all $t \in P$ computing $f \in F$, and $u
\sqsubseteq t$:

\[ \noo{S}{O}(u) \sqsubseteq f(\noo{S}{I}(u)) . \]

\end{lemma}

\proof\ Suppose firstly that $u$ is compact. Either $u = t$, in which case
the conclusion follows directly from the first condition for $t \in P$, or
by incrementality of $\Tee_S$, for some compact $v$, $\relcover{u}{v}{t}$.
Applying the second condition for $t \in P$,
\[ \noo{S}{O}(u) \sqsubseteq \noo{S}{O}(v) \sqsubseteq f(\noo{S}{I}(u)) . \]
The general result follows from this special case, since
\[ \noo{S}{O}(u) = \bigsqcup_{v \in K(u)} \noo{S}{O} (v) \sqsubseteq
\bigsqcup_{v \in K(u)} f(\noo{S}{I}(v)) = f(\noo{S}{I}(u)) . \quad \qed \]

\begin{lemma}

\label{lem2}

Let $\{ (S_j , P_j ) \}_{j \in J}$ be a non-deterministic functional network
computing $F_j$ at each $j \in J$, where $S_j = I_j \cup O_j$. Let $(S, P)
= \netcomp_{j \in J} (S_j , P_j )$. Then for all $t \in \Tee_S$:
\[ \begin{array}{rclr}
t \in P & \Leftrightarrow & \forall j \in J. \, \exists f_j \in F_j . & \\
& \bullet & \noo{S}{O_j}(t) = f_j (\noo{S}{I_j}(t)) & (1) \\
& \bullet & \relcover{u}{v}{t} \implies \noo{S}{O_j}(v) \sqsubseteq f_j
(\noo{S}{I_j}(u) & (2)
\end{array} \]

\end{lemma}

\proof\ We shall write $t_j = \restrict{S}{S_j}(t)$ for $t \in \Tee_S$. By
definition of network composition,
\[ \begin{array}{rclr}
t \in P & \Leftrightarrow & \forall j \in J. \, t_j \in P_j & \\
& \Leftrightarrow & \forall j \in J. \, \exists f_j \in F_j . & \\
& \bullet & \noo{S_j}{O_j}(t) = f_j (\noo{S_j}{I_j}(t_j )) & (1') \\
& \bullet & \relcover{u_j}{v_j}{t_j} \implies \noo{S_j}{O_j}(v)
\sqsubseteq f_j (\noo{S_j}{I_j}(u_j) & (2') 
\end{array} \]

Now it suffices to show that for all $t \in \Tee_S$, $j \in J$, $f_j \in
F_j$: $(1) \Longleftrightarrow (1')$ and $(2) \Longleftrightarrow (2')$. The
equivalence of $(1)$ and $(1')$ follows from the functoriality of $\rho$. To
show that $(2')$ implies $(2)$, we use the fact that $\rho$ weakly
preserves covers. Suppose $\relcover{u}{v}{t}$. If $u_j = v_j$, we can
apply Lemma~\ref{lem1} to get $(2)$; if $u_j \covers v_j$, we can apply
$(2')$. Finally, we show that $(2)$ implies $(2')$. Suppose
$\relcover{u'}{v'}{t_j}$. Since $\rho$ lifts covers, for some $u, v \in
\Tee_S$, 

\[ \restrict{S}{S_j}(u) = u' , \restrict{S}{S_j}(v) = v' , \AND
\relcover{u}{v}{t} . \]

We can now apply $(2)$ to get $(2')$, as required. \qed

As an immediate Corollary of Lemma~\ref{lem2}, we obtain:
\begin{proposition}
\label{globprop}
With notation as in Lemma~\ref{lem2}:
\[ \begin{array}{rclr}
t \in P & \Longleftrightarrow & \exists f \in \prod_{j \in J} F_j . & \\
& \bullet & \mu_S (t) = G_f (\mu_S (t)) & (1) \\
& \bullet & \relcover{u}{v}{t} \implies \mu_S (v) \sqsubseteq G_f (\mu_S (u))
& (2)
\end{array} \]
\end{proposition} 

{\sc Proof of Theorem~\ref{th1}}. With notation as in Lemma~\ref{lem2},
suppose $t \in P$. Applying Proposition~\ref{globprop} (1), for some $f \in \prod_{j \in J} F_j$,
$\mu_S (t) = G_f (\mu_S (t))$, whence $\lfp{G_f} \sqsubseteq \mu_S (t)$.
To show that $\mu_S (t) \sqsubseteq \lfp{G_f}$, let $( t_k )$ be a covering
sequence for $t$, which must exist by incrementality of $\Tee_S$; we
show by induction on $k$ that:
\[ \forall k \in \omega . (\mu_S (t_k ) \sqsubseteq G_f^k (\bot )). \]
The base case follows from the strictness of $\mu_S$. For the inductive
step,
\[ \begin{array}{rclrr}

\mu_S (t_{k + 1}) & \sqsubseteq & G_f (\mu_S (t_k )) & \mbox{Proposition
\ref{globprop} (2)} & \\

& \sqsubseteq & G_f (G_f^k (\bot )) & \mbox{by induction hypothesis.}
& \qed

\end{array} \]

\subsection{Liveness}

Consider an algebraic domain $D$, and a chain of compact elements $C =
(b_k )$ in $D$, with $\bigsqcup b_k = d$. We can consider $C$ as a (partial)
specification of a particular way of computing $d$, which induces a
causality relation on compact approximations of $d$, as follows. Define $
\| \cdot \|_C : K(d) \rightarrow \omega$ by
\[ \| b \|_C = \min \{ k \mid b \sqsubseteq b_k \} . \]

Now we can define:
\[ b <_C c \Iffdef \| b \|_C < \| c \|_C , \]
for $b, c \in K(d)$.

Now let $t$ be a trace in $\Tee_S$, with $\mu_S (t) = d \in \Vee_S$. We
can define a relation $<_t$ on $K(d)$ which reflects the causal constraints
on how $d$ can be realized introduced by $t$:
\[ \begin{array}{rcl}

b <_t c & \Iffdef & \mbox{for every covering sequence $(t_k )$ for $t$}: \\

& & \min \{ k \mid b \sqsubseteq \mu_S (t_k ) \} < \min \{ k \mid c
\sqsubseteq \mu_S (t_k ) \} .

\end{array} \]

\begin{definition}

Let $\Mod = (\Tee , \Vee , \mu )$ be an incremental model in which each
value domain $\Vee_S$ is $\omega$-algebraic. $\Mod$ is {\em causally
expressive} if for every sort $S$, $d \in \Vee_S$, and chain of compact
elements $C = (b_k )$ with $\bigsqcup b_k = d$, there exists $t \in
\Tee_S$ such that:

\begin{itemize}

\item $\mu_S (t) = d$

\item ${<_t} \supseteq {<_C}$.

\end{itemize}

\end{definition}

\begin{proposition}

\label{modprop2}

$\ModP$ and $\ModL$ are causally expressive.

\end{proposition}

\proof\ Since $\ModL$ is a sub-model of $\ModP$, it suffices to prove causal
expressiveness for $\ModL$. Suppose then that a compact chain $C = (b_n )$
in $\Ee_S$ is given, with $\bigsqcup b_n = d$. Since $\Ee_S$ is incremental,
$C$ can be refined into a covering sequence $C'$; clearly ${<_{C'}} \supseteq {<_C}$.
Now let $t$ be the trace in $\Ell{\Ee_S}$ corresponding to $C'$ under the
isomorphism of Proposition~\ref{covseqprop}. We note the general fact that 
for {\em any} algebraic cpo $D$, and covering sequence $(c_n )$ in $D$, there is
a {\em unique} covering sequence for $(c_n )$ in $\Cee{D}$; a consequence
of this is that $\Cee{\Cee{D}} \cong \Cee{D}$. If follows that ${<_t} = {<_{C'}} \supseteq {<_C}$, as required. \qed

We shall need a technical lemma about fixpoints in $\omega$-algebraic cpo's.
This was conjectured by the author, and proved under the hypothesis that
the domain is SFP. The ingenious proof of the general result is due to
Achim Jung (personal communication); it is reproduced here with his kind
permission.

\begin{lemma}[Jung]
\label{fixprop}
Let $D$ be an $\omega$-algebraic cpo, and $f : D \rightarrow D$ a continuous
function. There exists a chain $(b_n )$ of compact elements in $D$ such
that:
\begin{enumerate}
\item $b_0 = \bot$
\item $\forall n. \, b_{n+1} \sqsubseteq f(b_n )$
\item $\bigsqcup b_n = \lfp{f}$.
\end{enumerate}
\end{lemma}

\proof\ For each $f^n (\bot )$ we choose a chain of compact elements
$(c^n_m )$ with least upper bound $f^n (\bot )$. By taking a diagonal
sequence we find a chain $(c_n )$ with the property $c^{n'}_{m'} \sqsubseteq
c_n \sqsubseteq f^n (\bot )$ for all $n' , m' \leqslant n$. The least upper
bound of this chain is equal to $\lfp{f}$. Let $C_n = \diverges c_n$.

We shall define the required sequence $(b_n )$ inductively, to satisfy
the following properties:
\begin{enumerate}
\item $b_n \sqsubseteq f (b_{n - 1}), \quad n \geqslant 1$
\item $b_n \sqsubseteq f^n (\bot ), \quad n \geqslant 0$
\item $b_n \in O_n = \bigcap_{m \in \omega, 0 \leqslant 2m \leqslant n}
g^{-n + 2m} (C_{n - m}), \quad n \geqslant 0$.
\end{enumerate}
For $n = 2k$, the last property implies in particular that $b_n \in C_k$,
and together with (2) this ensures that the limit of the $b_n$ is the
least fixed point of $f$.

Let $b_0 = \bot$. Then (2) is obviously satisfied, and (3) evaluates to
\[ O_0 = f^0 (C_0 ) = C_0 = \diverges c_0 = \diverges \bot = D , \]
and is satisfied too.

Given $b_0 , \ldots , b_n$ we find $b_{n+1}$ as follows. First note that
$b_n \sqsubseteq f (b_{n-1}) \sqsubseteq f(b_n )$ by (1) (for $n = 0$
this is trivially satisfied); and that $f(b_n ) \sqsubseteq f^{n+1} (\bot )$
by (2). We shall select $b_{n+1}$ below $f(b_n )$ and above $b_n$, so
(1) and (2) will be satisfied. As for (3), we calculate:
\[ \begin{array}{l}
b_n \in O_n \implies f(b_n ) \in f(O_n ) \\
\subseteq \bigcap_{0 \leqslant 2m \leqslant n} f^{-n +2m +1} (C_{n-m}) \\
= \bigcap_{2 \leqslant 2m+2 \leqslant n+2} f^{-n-1+(2m+2)}(C_{n+1-(m+1)}) \\
= \bigcap_{2 \leqslant 2m' \leqslant n+2} f^{-n-1+2m'}(C_{n+1-m'}) \\
\subseteq \bigcap_{2 \leqslant 2m' \leqslant n+1} f^{-n-1+2m'}(C_{n+1-m'}).
\end{array}  \]
Note that $f^{n+1} (\bot )$ is contained in $C_{n+1}$, so we have
\[ \bot \in f^{-n-1}(f^{n+1}(\bot )) \subseteq f^{-n-1}(C_{n+1}), \]
which
tells us that $f^{-n-1}(C_{n+1}) = D$. So 
\[ f(b_n ) \in
\bigcap_{0 \leqslant 2m' \leqslant n+1} f^{-n-1+2m'}(C_{n+1-m'})
= O_{n+1} . \] 
Since $O_{n+1}$ is Scott-open, it contains a compact element below $f(b_n )$;
let $b_{n+1}$ be such an element above
$b_n$. \qed

\begin{theorem}

\label{th2}

If $\Mod$ is causally expressive, it satisfies $\GKPL$.

\end{theorem}

\proof\ We adopt the same notation as in Lemma~\ref{lem2}. Suppose $f
\in \prod_{j \in J} F_j$. We must show that for some $t \in P$, $\mu_S (t)
= \lfp{G_f}$. We apply Lemma~\ref{fixprop} to obtain a chain of
compact elements $C = (b_k )$ with $\bigsqcup b_k = \lfp{G_f}$, $b_0 =
\bot$, and $b_{k+1} \sqsubseteq G_f (b_k )$ for all $k$. Since $\Mod$ is
causally expressive, for some $t \in \Tee_S$, $\mu_S (t) = \bigsqcup b_k
= \lfp{G_f}$, and ${<_t} \supseteq {<_C}$. It remains to show that $t \in
P$. By Proposition~\ref{globprop}, it suffices to show that for all
$\relcover{u}{v}{t}$, $\mu_S (v) \sqsubseteq G_f (\mu_S (u))$, which in
turn is equivalent to:
\[ \forall b \in K(\Vee_S ). \, (b \sqsubseteq \mu_S (v) \implies b
\sqsubseteq G_f (\mu_S (u)) ). \]
Suppose then that $b \sqsubseteq \mu_S (v) \sqsubseteq \mu_S (t) =
\lfp{G_f}$. Since $b$ is compact, $b \sqsubseteq b_k$ for some $k$. If $b =
\bot$ we are done; otherwise, for some $k$, $b \sqsubseteq b_{k+1}$, $b
\not\sqsubseteq b_k$. This implies $b_k <_C b$, and hence $b_k <_t b$. By
incrementality of $\Tee_S$, we can find a covering sequence $(t_k )$ for
$t$ with $u = t_n$, $v = t_{n+1}$ for some $n$. But since $b \sqsubseteq
\mu_S (v)$ and $b_k <_t b$, this implies $b_k \sqsubseteq \mu_S (u)$, and
hence
\[ b \sqsubseteq b_{k+1} \sqsubseteq G_f (b_k ) \sqsubseteq G_f (\mu_S
(u)) , \]
as required. \qed

As an immediate Corollary of Propositions~\ref{modprop1} and
\ref{modprop2} and Theorems~\ref{th1} and \ref{th2}, we obtain:

\begin{theorem}

$\ModP$ and $\ModL$ satisfy (GKP).

\end{theorem}

\section{Concluding Remarks}

The results in this paper are of a preliminary nature. Even within the
asynchronous network model, there are a number of interesting topics for
further investigation. These include the characterisation of models in
terms of properties of {\em extensionality} and {\em expressive
completeness}; and connections with {\em full abstraction}. Also, it would
be of interest to specify a uniform operational semantics for our general
class of models $\ModP$, and to prove some correspondence results.
A good basis for this should be given by \cite{Cur86a}.
It would also be interesting to formulate a notion of {\em continuous}
({\em e.g.} probabilistic) computation in a network, replacing
algebraic domains by continuous ones. Much of the theory developed here
should generalize; note in particular that Lemma~\ref{fixprop} is valid
for $\omega$-continuous cpo's, replacing ``compact'' by ``relatively
compact''.
Beyond asynchronous networks, we would like to give a general notion of
model in categorical terms, which would subsume a wide range of
concurrency formalisms, including process algebras and Petri nets, as well as dataflow. The ideas of \cite{Win88} should be relevant here.

{\bf Acknowledgements.} I would like to thank Jay Misra for providing
the initial stimulus to this work by sending me his paper \cite{Mis89};
much of the present paper can be seen as an attempt to understand some of
his ideas in a general setting.
I would also like to thank Achim Jung for helpful discussions while the ideas developed, and
for proving Lemma~\ref{fixprop}; and Mike Mislove for inviting me to
the 1989 MFPS conference, where I presented this material at an informal
``pre-meeting''; and for inviting me to submit this paper to the
Proceedings. My thanks also to my hosts at the University of Pennsylvania
for providing such a stimulating and friendly environment during my
visit in the first half of 1989; the Nuffield Foundation, for their
support in the form of a Science Research Fellowship for 1988--89; and
the U.K. SERC and U.S.A. NSF for their financial support.
\bibliography{biblio,dfbib}

\end{document}